# Decomposition Process of Carboxylate MOF HKUST-1 Unveiled at the Atomic Scale Level

AUTHOR NAMES

Michela Todaro, [a,b] Gianpiero Buscarino, [a,*] Luisa Sciortino, [a] Antonino Alessi, [a,c] Fabrizio Messina, [a] Marco Taddei, [d] Marco Ranocchiari, [d] Marco Cannas, [a] Franco M. Gelardi [a]

AUTHOR ADDRESS

[a] Dipartimento di Fisica e Chimica, Università di Palermo, 90123 Palermo, Italia;

[b] Dipartimento di Fisica e Astronomia, Università di Catania, 95123 Catania, Italia;

[c] Laboratoire H. Curien, UMR CNRS 5516, Université de Lyon, 42000 Saint-Etienne, France;

[d] Paul Scherrer Institute, Laboratory for Catalysis and Sustainable Chemistry (LSK) 5232 Villigen, Switzerland.

ABSTRACT

HKUST-1 is a metal-organic framework (MOF) which plays a significant role both in applicative and basic fields of research, thanks to its outstanding properties of adsorption and catalysis but also because it is a reference material for the study of many general properties of MOFs. Its metallic group comprises a pair of $Cu^{2+}$ ions chelated by four carboxylate bridges, forming a structure known as paddle-wheel unit, which is the heart of the material. However, previous studies have well established that the paddle-wheel is incline to hydrolysis. In fact, the prolonged exposure of the material to moisture promotes the hydrolysis of Cu-O bonds in the paddle-wheels, so breaking the crystalline network. The main objective of the present experimental investigation is the determination of the details of the structural defects induced by this process in the crystal and it has been successfully pursued by coupling the electron paramagnetic resonance spectroscopy with other more commonly considered techniques, as X-ray diffraction, surface area estimation and scanning electron microscopy. Thanks to this original approach we have recognized three stages of the process of decomposition of HKUST-1 and we have unveiled the details of the corresponding equilibrium structures of the paddle-wheels at the atomic scale level.

INTRODUCTION

Metal-organic frameworks (MOFs) are porous hybrid compounds constituted by organic molecules and metal ions linked together to form crystalline networks with high surface area and large pore volume.[1-4] The family of MOFs is very numerous due to the large number of metal ions/groups and organic linkers which can be combined to give stable crystalline structures by appropriate synthesis. The obtained structures have a large variety of microporous crystalline frameworks and physicochemical properties, resulting of potential interest for a wide range of different applications.[1]

One of the most promising MOF is HKUST-1,[5] whose chemical formula is $Cu_3(BTC)_2(H_2O)_3$. Thanks to the great affinity of its metallic group with $NH_3$ and $CO_2$ molecules, it can be used to remove such toxic gases, ammonia[6-7] and carbon dioxide,[8-9] from contaminated air and flue gases. In energetic and environmental fields, HKUST-1 can be used for separation of carbon dioxide from hydrogen, methane, nitrogen, and oxygen to purify these gases and to decrease the emission of greenhouse gases.[9-11] Its potentialities for gas storage, in particular for hydrogen and methane, are relevant in the field of alternative and renewable energies.[12-16] As catalyst, HKUST-1 is very promising because its metal sites are strongly active in many reactions and because they can be easily accessed from the surface of the large pores of the framework by a large variety of interesting molecules.[17] Recently, thin films of HKUST-1 on opportune substrates have been successfully used in photovoltaic applications.[18] Finally, it is worth noting that HKUST-1 is a material of particular interest also because it has gained in a short time the role of reference system for the study of a lot of general properties of MOFs and, in particular, of carboxylate MOFs.[1-4]

The metallic group of HKUST-1 involves a pair of $Cu^{2+}$ ions (each with electron spin S=1/2) coordinated by four carboxylate bridges to form a paddle-wheel moiety; each carboxylate group is part of a benzene 1,3,5-tricarboxylate (BTC) linker molecule (Figure 1).[5,9] In the as synthetized HKUST-1, the fifth (out of plane) binding site on each $Cu^{2+}$ ion is occupied by the oxygen of a crystallization water molecule which can be easily removed by the activation process (typically consisting in a thermal treatment in vacuum at about 400 K).[19]

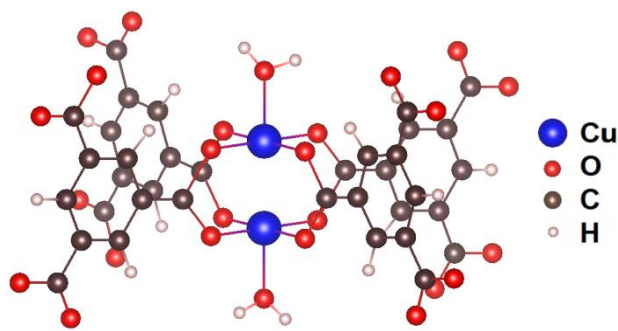

**Figure 1:** Sketch of the main structure of HKUST-1 before activation. It shows the paddle-wheel structure which is formed by bonding two $Cu^{2+}$ ions with four benzene 1,3,5-tricarboxylate (BTC) linker molecules.

Studies on magnetic properties of HKUST-1 have shown that the two S=1/2 spins of copper ions are weakly coupled through a super-exchange interaction, which involves wave functions delocalized on the carboxylate bridges.[20-23] This interaction leads to an antiferromagnetic coupling, which determines a diamagnetic ground state with total spin S=0 and a paramagnetic excited state with total spin S=1.[20-23] The strength of the super-exchange coupling is essentially related to the high symmetry of the paddle-wheels. As a consequence, even the adsorption processes of small molecules, as $H_2O$ or $NH_3$, on the axial binding site of $Cu^{2+}$ are able to reduce significantly this coupling.[6,19] Larger effects are induced on the magnetic coupling when a change of the coordination number of $Cu^{2+}$ ion takes place.[24]

Previous studies on water uptake have pointed out that activated HKUST-1 is strongly hydrophilic[25] and that the open metal sites on $Cu^{2+}$ ions are the primary adsorption sites for polar molecules such as water.[25-27] For example, few minutes of exposure to air are sufficient to reveal water uptake with NMR spectroscopy.[28] This affinity of HKUST-1 with respect to water is very interesting for many applications, as in biomedical field,[29-30] in the packing and transporting of gaseous industrial streams to avoid corrosive and condensation effects[31] and to increase the efficiency of $CO_2$ capture.[32]

In spite of these interesting applications, it has been shown that the crystalline structure of the HKUST-1 can be seriously damaged by prolonged exposure to moist environment. In fact, in these conditions water molecules are able to condensate into the cavities of the material and finally it becomes energetically favorable for $H_2O$ to hydrolyze the Cu-O bond involved in the paddle-wheel, so breaking the crystalline structure.[25-26,33-35] This property limits not only the applications where the HKUST-1 is directly exposed to moisture or water, but also all those uses where water is present just as impurity.[3,36-38] Even though the deleterious effects related to hydrolysis play such important role for HKUST-1 and for the whole class of the carboxylate MOFs, the stages of the decomposition process induced by exposure to moisture have never been identified and characterized in detail, yet.

Here we present an experimental investigation focused on the process of structural degradation induced in HKUST-1 by exposition to moisture. Our study has involved different spectroscopic/microscopic techniques: X-ray diffraction (XRD), Nitrogen isotherms for surface area determination with Brunauer–Emmett–Teller (BET) method, scanning electron microscopy (SEM) and electron paramagnetic resonance (EPR) spectroscopy. In particular, this latter method has played a key role in the present work because it is very sensitive to the magnetic properties of the paddle-wheels, which in turn are very sensitive to the local structures around the magnetic $Cu^{2+}$ ions. This original approach, thanks to the well-known potentiality of the EPR technique in the determination of the atomic scale structure of the paramagnetic centers investigated,[39-40] has permitted us to point out for the first time that the decomposition process of HKUST-1 actually takes place through three different stages. These stages are characterized by different equilibrium configurations of the paddle-wheels, whose structures are here presented and discussed in detail.

MATERIALS AND METHODS

HKUST-1 was purchased from Sigma-Aldrich as Basolite C300 in powder form and it was firstly used to obtain two samples, each with mass of about 70 mg. The first sample, hereafter named *SAMPLE A*, was activated in an EPR tube under vacuum at 400 K for 8 h. To preserve this sample from the interaction with the moisture of air, the tube was sealed with a blowtorch just after the activation process. This sample was found to maintain its typical navy blue color. The second sample, hereafter named *SAMPLE B*, was continuously exposed to air at 300 K and 70% relative humidity (RH) for different times up to 200 days. This latter sample has totally changed its color from navy blue to turquoise in few minutes. Both these samples were monitored by EPR, for comparison. In addition to the *SAMPLE A* and *B* described above, many other samples of type *B* were prepared in similar way and they were used for XRD and SEM measurements, for BET analysis and for the reactivation experiments.

XRD measurements were carried out using a Bruker *D5000* diffractometer in Bragg Brentano geometry ($\lambda_{Cu\ K\alpha}$=0.154 nm) at 300 K and by using an $SiO_2$ zero-diffraction plate. The XRD data were collected in the angular range of 4÷60° in 2θ using a 0.01° step size.

SEM analyses were carried out using an SEM-Oxford *LEO 440* with an acceleration voltage of 4 kV. A coating with graphite by an EMITECH K950X graphite evaporator was made before SEM acquisitions.

Nitrogen adsorption isotherms were recorded on a Micromeritics *TriStar II 3020 V1.03* at 77 K. Surface areas were determined using the BET method in a relative pressure range of $P/P_0$=0.009 to 0.026. Before measurements all the samples were activated at 400 K under vacuum overnight, using the same sample cell.

EPR measurements were carried out using a Bruker *EMX micro* spectrometer working at a frequency of about 9.5 GHz (X-band). The magnetic-field modulation frequency was 100 kHz. The spectra were acquired by putting the glass tube containing the sample into a dewar flask. This latter was filled with liquid nitrogen for measurements at 77 K, whereas it was empty (unfilled with liquid nitrogen) for measurements at 300 K. The absolute concentrations of the paramagnetic centers in the samples considered were estimated by comparing the double integrals of the EPR lines with that of a reference sample. The defects concentration in the reference sample was evaluated, with absolute accuracy of 20%, using the instantaneous diffusion method in spin-echo decay measurements carried out in a pulsed EPR spectrometer.[41] Spectral simulations of the CW EPR powder patterns were done using the *EasySpin ESR simulation package*.[42]

RESULTS

### *SAMPLE A*

EPR spectra of *SAMPLE A* acquired at different times after its preparation and obtained at 77 K are reported in Figure 2.

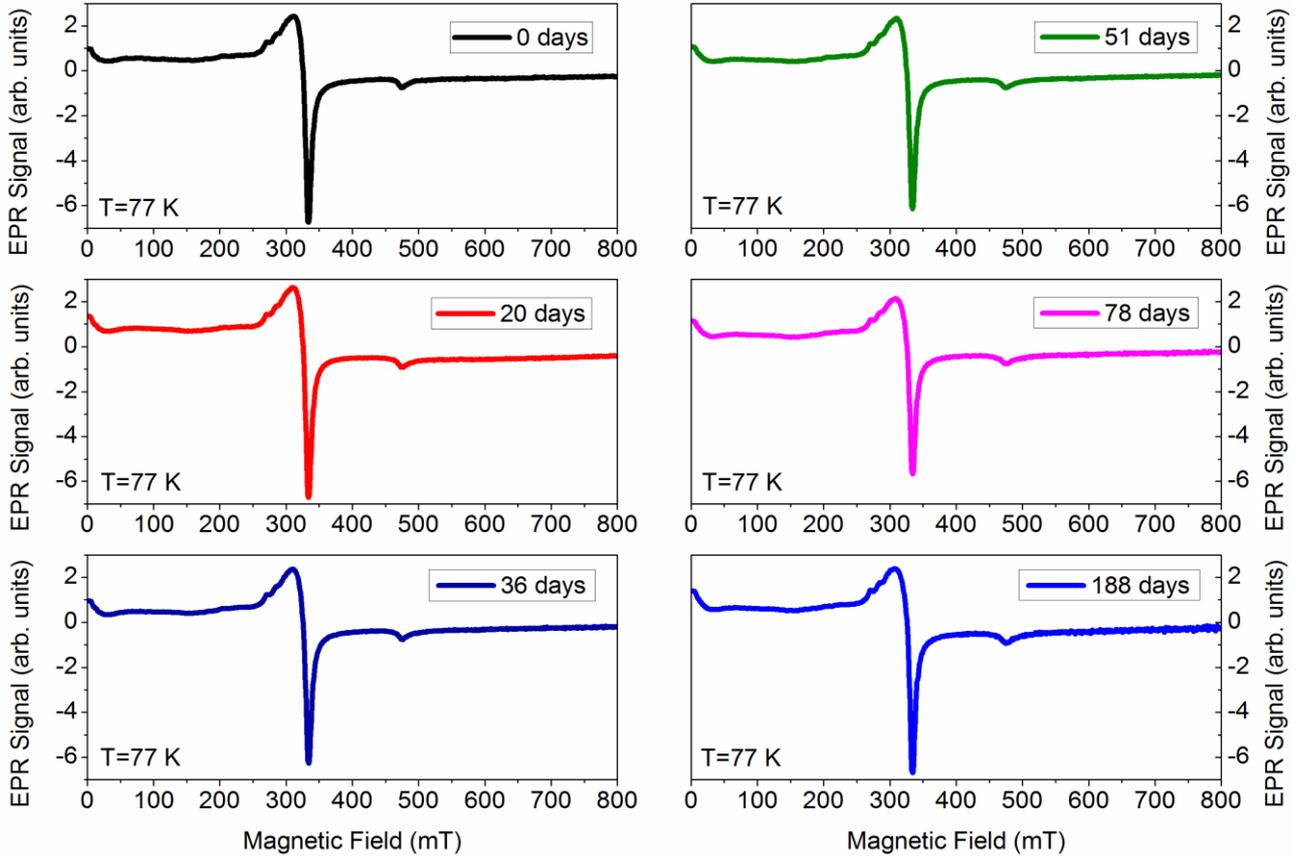

**Figure 2:** EPR spectra of the *SAMPLE A* of HKUST-1 acquired at different times after its preparation and obtained at 77 K.

All the EPR spectra of *SAMPLE A* shown in Figure 2 have common spectroscopic characteristics: (i) a main resonance centered at $B_0=325$ mT and peak-to-peak line width of about $\Delta B_{pp}=24$ mT, on which traces of an hyperfine structure are barely evident, and (ii) two smaller contributions at about $B_0=12$ mT and $B_0=470$ mT. In accordance with literature,[20,43-45] we assign the central resonance peaked at $B_0=325$ mT to the complex $[Cu(OH_2)_6]^{2+}$, which involves a $Cu^{2+}$ ion, with electron spin $S=1/2$, coordinated with six water molecules and not connected with the framework of HKUST-1. It presumably is a defect formed during the material synthesis.[20] The hyperfine structure, composed by four lines, is due to the interaction of the $Cu^{2+}$ electron spin $S=1/2$ with its nuclear spin $I^{Cu}=3/2$.

The two smaller peaks at about $B_0=12$ mT and $B_0=470$ mT observed in the EPR spectra acquired at 77 K are due to a randomly oriented triplet centers with axial symmetry.[20,43-44,46-48] This paramagnetic center with $S=1$ electron spin, hereafter named E"(Cu),[49] originates from the antiferromagnetic coupling between the two spin $S=1/2$ of the couple of $Cu^{2+}$ ions involved in each paddle-wheel.[20,43-44,46-48]

EPR spectra of *SAMPLE A* acquired at different times after its preparation and obtained at 300 K are reported and discussed in Figure S1. XRD, SEM and BET measurements were not possible for *SAMPLE A* because, as discussed above, it was sealed in a glass tube just after activation. However, it is widely reliable to assume that this sample preserves its pristine crystalline structure over the time basing on the fact we have observed no changes of its EPR spectrum nor of its color. In addition, we have verified that both these features actually coincide with those attributed in literature to fresh activated HKUST-1.[6,20]

## *SAMPLE B*

XRD patterns obtained for *SAMPLE B* exposed to air (300 K, 70% RH) for 1, 30 and 165 days are reported in Figure 3.

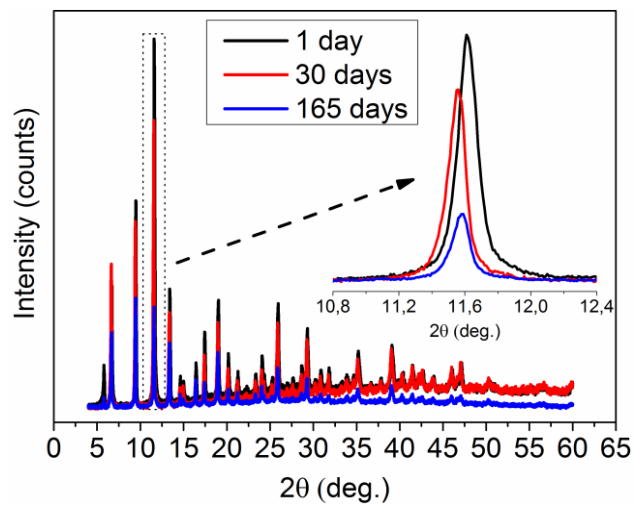

**Figure 3:** XRD patterns of the *SAMPLE B* of HKUST-1 exposed to air (300 K, 70% RH) for 1 day (black curve), 30 days (red curve) and 165 days (blue curve). The inset shows a zoom of the most intense line peaked at about 2θ=11.6 deg.

The diffraction pattern obtained during the 1 day of exposure to air, black line in Figure 3, clearly shows the characteristic peaks of HKUST-1,[8-9,17,21,25-26,33,38,50-51] confirming the high quality of the starting crystalline material. After 30 days of exposure to air, red line in Figure 3, we note a reduction of the intensity of the peaks and their systematic shift to smaller angles. These latter minor changes are recognizable in the inset of Figure 3, where we report a zoom of the most intense peak of the pattern. The blue line in Figure 3 is the diffraction pattern obtained for *SAMPLE B* after exposure to air for 165 days. In this case we observe a further reduction of peaks intensities and a peaks shift back to higher angles.

SEM images obtained for *SAMPLE B* are reported in Figure 4, where it is possible to compare the morphology of HKUST-1 crystals after different times of exposure to air (300 K, 70% RH).

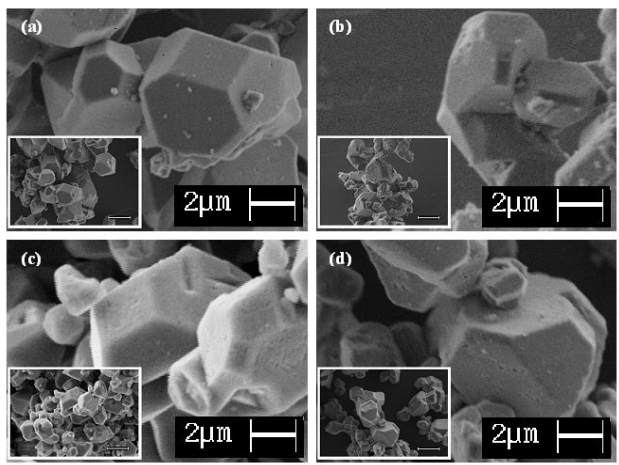

**Figure 4:** SEM images of the *SAMPLE B* of HKUST-1 exposed to air (300 K, 70% RH) for different times: (a) 0 days, (b) 20 days, (c) 36 days and (d) 188 days.

In particular, Figure 4a shows that the grains of the fresh sample appear well recognizable and regular crystals with sharp edges, according to other reports in literature.[8-9,38] After 20 days of exposure to air, Figure 4b, no evident morphological changes are observed. At variance, in the samples exposed to air for 36 days and 188 days, Figures 4c and 4d, respectively, we recognize a more notable appearance of a lot of small holes on their surfaces. All SEM images reported in Figure 4, are selected from a large set of images and demonstrative of the average state in which the sample was.

Surface area and pore volume were estimated for *SAMPLE B* after different times of exposure to air (300 K, 70% RH) by fitting adsorption curves of $N_2$ at 77 K (reported in Figure S2) with BET equation.[52] The results of these fit are summarized in Table 1. As shown, a regular behavior is observed, as pores volumes change consistently with BET surface area.

**Table 1.** Values of BET surface area and micropore volume of HKUST-1 at different time of exposure to air.

| time of exposure to air (days) | BET surface area (m$^2$/g) | micropore volume (cm$^3$/g) |
|:---:|:---:|:---:|
| 0 | 1716 | 0.6 |
| 20 | 1813 | 0.6 |
| 36 | 960 | 0.3 |
| 188 | 1097 | 0.4 |

The BET surface area, estimated for fresh HKUST-1 (0 days), is in line with those reported in literature for this material.[8,12,25,53-57] Interestingly, in the sample exposed to air for 20 days, no degradation of the natural porosity of the material is observed, suggesting that the network preserves the pristine structure. This is also evidenced by the shape of the adsorption isotherm of this sample, which is an ideal Type I curve, as observed for the fresh sample (Figure S2). At variance, after 36 days and 188 days of exposure to air, BET surfaces and pore volumes are both reduced by about 50% with respect to the fresh sample. Furthermore, the adsorption isotherms display a significant deviation from the ideal Type I shape (Figure S2). A similar effect has been reported in previous works[25-26,33] and indicates that the network has been irreversibly damaged by prolonged interaction with air moisture.

EPR spectra acquired for *SAMPLE B* at 77 K after exposure to air (300 K, 70% RH) for different times are shown in Figure 5.

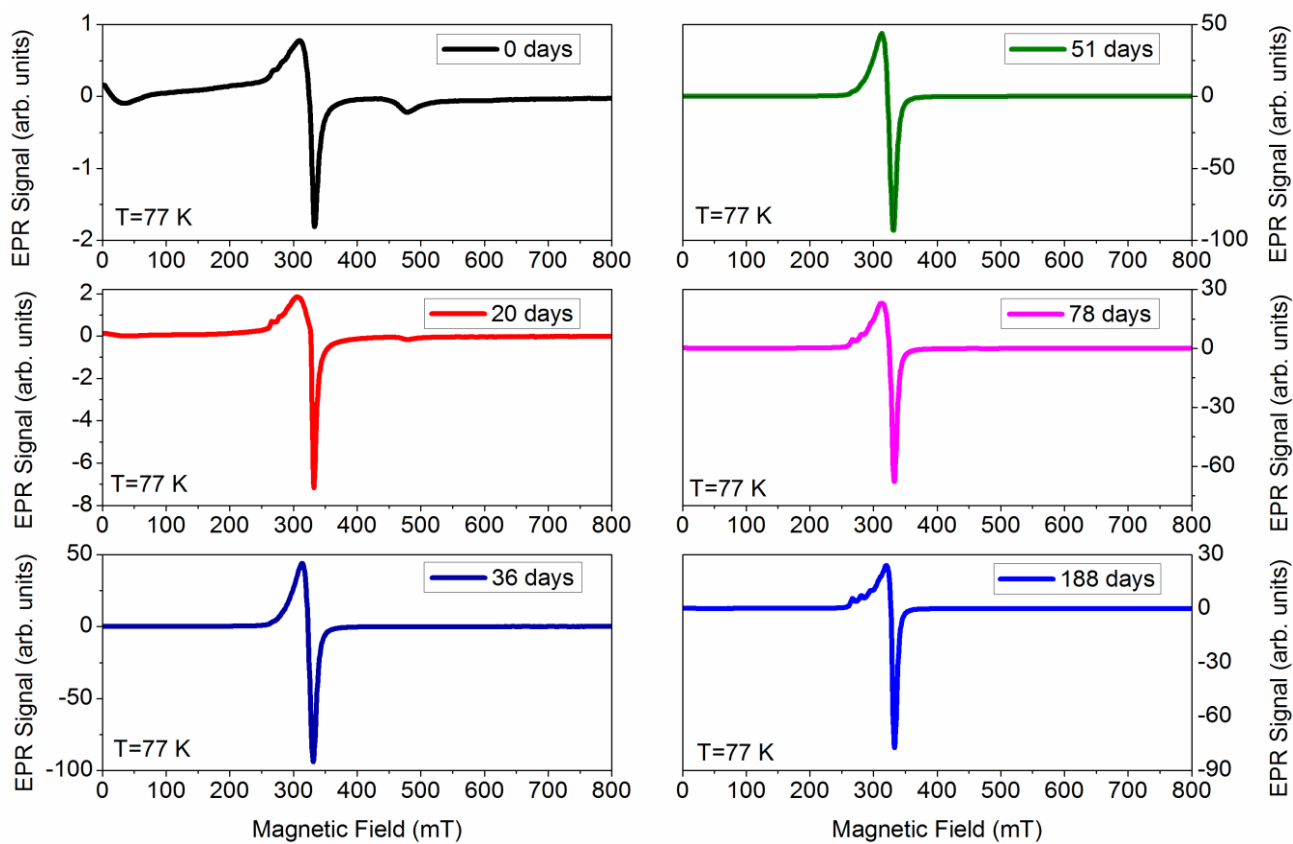

**Figure 5:** EPR spectra of the *SAMPLE B* of HKUST-1 acquired at different times after its preparation and obtained at the temperatures of 77 K.

These measurements indicate that relevant changes of the EPR spectra take place when HKUST-1 is exposed to air. In fact, while for the fresh sample (0 days) we observe just the same EPR lines found in the *SAMPLE A* (Figure 2), upon exposure to air moisture these signals gradually disapper (vide infra) leaving the place to two new resonances with quite different spectroscopic properties. These latter lines were observed virtually isolated from other contributions after exposure of HKUST-1 to air moisture for 36 and 188 days, respectively (Figure 5). Interestingly, we have found that the ratio of the EPR signal intensities obtained for each of these latter two new resonances at 77 K and at 300 K (these latter measurements are reported in Figure S3) is about four, as prescribed by Curie law. This result unveils that they actually involve simple S=1/2 paramagnetic species and not pairs of weakly coupled electrons, as is the case for the two $Cu^{2+}$ ions in the pristine paddle-wheels.

DISCUSSION

In order to understand in deep the processes taking place in HKUST-1 upon exposure to moisture, we have performed a detailed study of the EPR spectra obtained for the *SAMPLE B* after different times of exposure to air (Figure 5 and Figure S3). Thanks to this comprehensive analysis, we have recognized that all the spectra at fixed temperature result from the superposition, with different weights, of just four components lines which are attributable to four distinguishable paramagnetic centers. In particular, the first of them observed in the fresh *SAMPLE B*, obviously consists in the well-known E"(Cu) center, originating from the magnetic coupling of the two S=1/2 located in the $Cu^{2+}$ ions involved in the paddle-wheels.[20] In the EPR spectra obtained at 77 K this contribution usually superimposes to that attributed to the extraframework complex $[Cu(OH_2)_6]^{2+}$, as discussed above, which represents the second species we have recognized. The latter two paramagnetic centers we have isolated, were observed in *SAMPLE B* after exposure to air for 36 and 188 days and hereafter will be referred to as $E'_1(Cu)$ and $E'_2(Cu)$,[49] respectively. As anticipated, they exhibit quite different nature with respect to the E"(Cu) center, as they consist in simple S=1/2 paramagnetic species. It is worth noting that EPR lines somewhat similar to those we attribute to these centers have already been reported in previous experimental works devoted to HKUST-1 exposed to moisture. They were tentatively associated to some Cu-related decomposition products of the material.[21,58] Nevertheless, the problem of the attribution of these EPR signals to specific atomic scale structures was not investigated in detail.

In order to better understand the properties of all the characteristic paramagnetic centers observed in the *SAMPLE B* after exposure to air for different times, we have performed computer simulations of their EPR spectra acquired at 77 K by using the procedure detailed in the following. Since the EPR spectra of the E"(Cu) center and of the $[Cu(OH_2)_6]^{2+}$ complex were already successfully characterized in detail in a previous work,[20] for our simulations we have chosen to consider the same Hamiltonians models used in that work. They are reported in Eq. (1) and (2), respectively.

$$H = \beta_e \mathbf{B} \cdot \mathbf{g} \cdot \hat{\mathbf{S}}_{TOT} + \hat{\mathbf{S}}_{TOT} \cdot \mathbf{D}_{dip} \cdot \hat{\mathbf{S}}_{TOT} + J_0 \left( \frac{1}{2} \hat{S}^2_{TOT} - \frac{3}{4} \mathbf{1} \right) \quad (1)$$

$$H = \beta_e \mathbf{B} \cdot \mathbf{g} \cdot \hat{\mathbf{S}} + \hat{\mathbf{S}} \cdot \mathbf{A}^{Cu} \cdot \hat{\mathbf{I}}^{Cu} \quad (2)$$

The Hamiltonian of Eq. (1) is written for a paramagnetic center with the structure shown in Figure 1, which includes two S=1/2 centers related to the two $Cu^{2+}$ ions involved in the paddle-wheel. As discussed above, a property characterizing such structures is that the two S=1/2 centers actually establish a measurable antiferromagnetic coupling through the wave function delocalized over the four carboxylate bridges, which chelate the two $Cu^{2+}$ ions. At variance, the Hamiltonian of Eq. (2) is written for a paramagnetic center, which simply involves a single S=1/2 center related to the just a single $Cu^{2+}$ ion. In this latter case no couple of interacting S=1/2 centers are involved. More in details, in Eq. (1) and (2), $\beta_e$ is the Bohr magneton, **B** is the external magnetic field, **g** is the spectroscopic splitting matrix and $\mathbf{S}_{TOT} = \mathbf{S}_1 + \mathbf{S}_2$ is the total spin operator obtained summing spin operator of the two $Cu^{2+}$ ions. The first term in Eq. (1) describes the *Zeeman Hamiltonian*, the second describes the *electron dipole-dipole interaction* through the $\mathbf{D}_{dip}$ operator and the third term the *exchange interaction* through the isotropic electron-exchange coupling constant $J_0$. The $\mathbf{D}_{dip}$ matrix, with trace zero, is usually reported in terms of the parameters D and E. These latter two parameters are defined as $D = 3/2 \cdot D_{zz}$ and $E = 1/2 \cdot (D_{xx} - D_{yy})$, where $D_{xx}$, $D_{yy}$ and $D_{zz}$, are the principal elements of the $\mathbf{D}_{dip}$ matrix. In the case of randomly oriented triplet center with axial symmetry, $D \neq 0$ and $E \approx 0$.[39-40,47] The first term in Eq. (2) is again the *Zeeman Hamiltonian*, whereas the second term describes the *hyperfine interaction* between the unpaired electron spin and the nuclear spin of the $Cu^{2+}$ ion. Copper has two magnetic isotopes with the following properties: (i) $^{63}Cu$, natural abundance 69.17%, isotropic hyperfine parameter 213.92 mT and (ii) $^{65}Cu$, natural abundance 30.83%, isotropic hyperfine parameter 228.92 mT.

In contrast to the E"(Cu) center and $[Cu(OH_2)_6]^{2+}$ complex, the E'$_1$(Cu) and E'$_2$(Cu) centers have never been characterized in detail before, so no reference Hamiltonians models can be found in literature for them. However, since the spectroscopic features of their EPR lines indicate they both reasonably involve a single unpaired spin on a Cu ion, for these latter centers we have considered the same Hamiltonian model used for the $[Cu(OH_2)_6]^{2+}$ complex reported in Eq. (2).

The optimized EPR lines obtained by computer simulations using the Hamiltonian of Eq. (1) and (2) are shown in Figure 6, whereas all the relative optimized parameters are collected in Table 2. In Figure 6, the well-known EPR line pertaining to $[Cu(OH_2)_6]^{2+}$ complex[20] has been omitted, for simplicity. The comparison between the experimental and the simulated spectra are presented in Figures S4-S6.

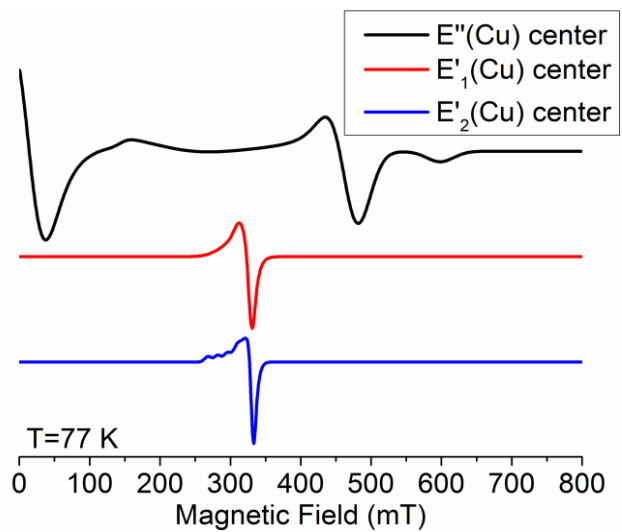

**Figure 6:** Optimized EPR lines obtained by computer simulation for the E"(Cu), E'$_1$(Cu), E'$_2$(Cu) centers observed in the *SAMPLE B* at T=77 K.

In Table 2 are reported principal values of **g** matrix and their strains, principal values of **A** matrix and D and E parameters pertaining to **D**$_{dip}$ matrix. The introduction of **g** strain was found to be necessary to obtain good optimized EPR lineshapes and it corresponds to assume that the principal **g** values of the paramagnetic centers follows a Gaussian distributions whose widths are given by the corresponding values of the **g** strains.

**Table 2. Optimized Hamiltonian parameters resulting from fit procedure of simulated EPR spectra reported in Figure 6.**

| paramagnetic center | $g_1$ | $g_2$ | $g_3$ | $g_1$ strain | $g_2$ strain | $g_3$ strain | $A_1$ (cm$^{-1}$) | $A_2$ (cm$^{-1}$) | $A_3$ (cm$^{-1}$) | D (cm$^{-1}$) | E (cm$^{-1}$) |
|---|---|---|---|---|---|---|---|---|---|---|---|
| E"(Cu) | 2.07$^a$ | 2.07$^a$ | 2.32 | 0.28 | 0.28 | 0.36 | - | - | - | 0.330 | 0 |
| E'$_1$(Cu) | 2.06 | 2.12 | 2.24 | 0.03 | 0.12 | 0.27 | 1.8x10$^{-3}$ | 4x10$^{-4}$ | 1.3x10$^{-2}$ | - | - |
| E'$_2$(Cu) | 2.08 | 2.09 | 2.37 | 0.03 | 0.12 | 0.09 | 1x10$^{-3}$ | 1.4x10$^{-3}$ | 1.5x10$^{-2}$ | - | - |
| [Cu(OH$_2$)$_6$]$^{2+}$ | 2.08 | 2.07 | 2.35 | 0.25 | 0.04 | 0.14 | 1x10$^{-3}$ | 1.4x10$^{-3}$ | 1.4x10$^{-2}$ | - | - |

*a* In line with previous works,[20] the axial symmetry has been assumed for the E"(Cu) center before optimization of the parameters.

The EPR spectra obtained for the *SAMPLE B* after exposition to air for different times were all analyzed as a linear combination of the references ones, attributed to the E"(Cu), E'$_1$(Cu), E'$_2$(Cu) centers and the $[Cu(OH_2)_6]^{2+}$ complex. The coefficients of these linear combinations were obtained by a fit procedure and have permitted us to estimate the concentrations of the component centers in all the experimental spectra of interest. These data are collected in Figure 7 where we report the concentration of E"(Cu), E'$_1$(Cu), E'$_2$(Cu) centers in the *SAMPLE B* after exposure to air for different times. The concentration of $[Cu(OH_2)_6]^{2+}$ complex remains almost constant at the value of about $2\times10^{19}$ spin/cm$^3$ and it has been omitted in Figure 7, for clarity.

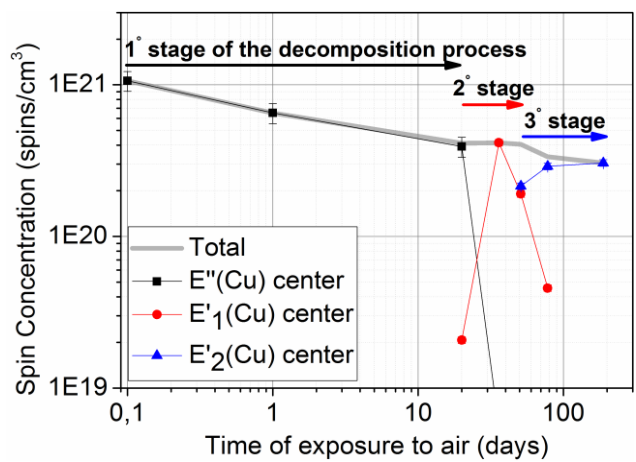

**Figure 7:** Spin concentration of E"(Cu), E'$_1$(Cu) and E'$_2$(Cu) paramagnetic centers in *SAMPLE B* of HKUST-1 as a function of time of exposure to air (300 K, 70% RH).

As shown, the main effect observed in the first 20 days of exposure to air is the reduction of the concentration of the E"(Cu) centers. Nevertheless, in the spectrum obtained after 20 days a barely detectable contribution arising from the E'$_1$(Cu) is observed, which is found to increase significantly for longer exposures up to 36 days. It is worth noting that the maximum concentration of E'$_1$(Cu) centers induced in the sample is almost equal to the concentration of residual E"(Cu) centers observed after 20 days of exposure, suggesting that the former center may be generated from the latter as a consequence of some local structural modification induced by prolonged exposition to air. Upon further exposure to air the E'$_1$(Cu) centers gradually anneal out leaving the place to the E'$_2$(Cu) centers after about 50 days.

From a general point of view, the overall data reported above, and particularly those in Figure 7, put in evidence the existence of three different stages of the process of decomposition of HKUST-1 upon exposure to air which, at the temperature and relative humidity considered in the present work, are observed for the following exposure times, $\tau$, ranges: (i) $0 < \tau < 20$ days, (ii) $20 < \tau < 50$ days and (iii) $\tau > 50$ days. These stages are discussed in detail in the following.

As shown in Figure 7, the main effect observed by EPR in the first stage of the process ($0 < \tau < 20$ days) is a reduction of the concentration of E"(Cu) centers of about 65%. This property, recognized for the first time in the present work, is very important because it proves that some relevant structural changes affect the majority of the paddle-wheels involved in the material. In order to understand if these structural variations actually affect the stability of the network of the HKUST-1 or not, it is instructive to compare BET surfaces and pore volumes estimated in the fresh material with those obtained after 20 days of exposure to air (Table 1). As shown in Table 1, comparable values are obtained for HKUST-1 before and after exposure to air for 20 days, proving that the processes responsible for the decrease of the concentration of E"(Cu) centers do not modify irreversibly the network of the material. In agreement with this conclusion, we have observed that the pristine concentration of E"(Cu) centers is actually recovered by reactivation of the sample (Figure S7). All these data suggest that this first stage of the interaction of the HKUST-1 with moisture does not involve hydrolysis, which would cause irreversible modifications. In line with this picture, SEM images do not show any significant change in the morphology of HKUST-1 after 20 days of exposure to air with respect to the pristine material (compare Figures 4a and 4b). A simple and consistent way to understand the experimental evidences collected for HKUST-1 during the first 20 days of exposure to air, is that to suppose that during this time many couples of $Cu^{2+}$ ions, pertaining in the same paddle-wheels, are forced to establish a σ bond between them. The formation of this bond does not damage irreversibly the network, but it makes the related couples of unpaired electrons, which constitute the E"(Cu) centers, to become paired and consequently to lose their EPR signal, in agreement with the experimental observations. This structural change is shown in Figure 8.

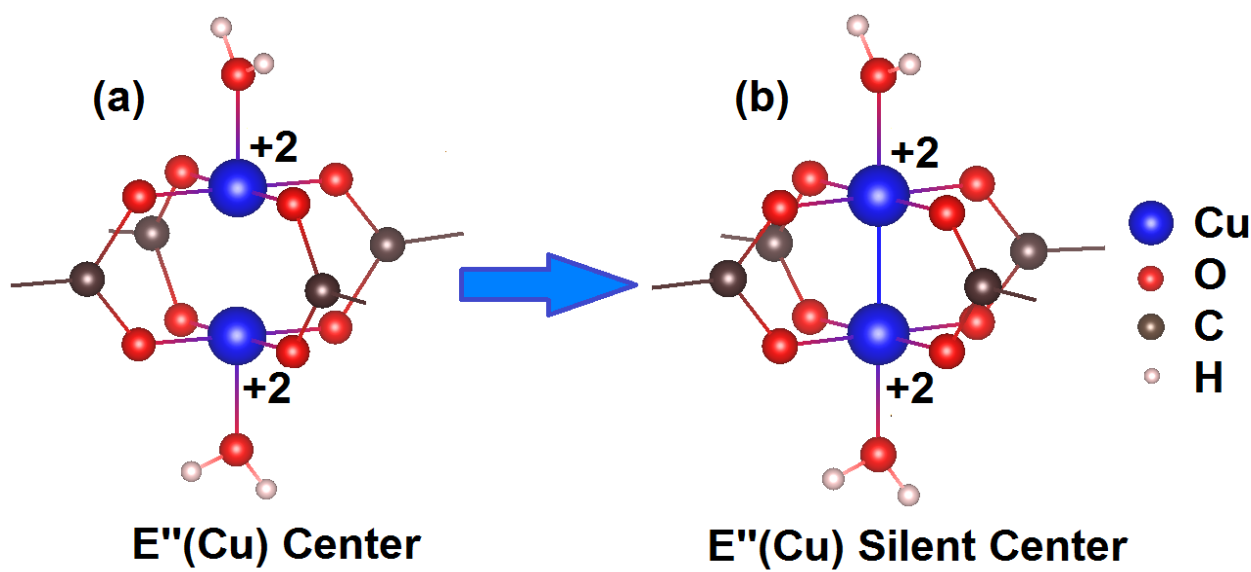

**Figure 8:** Schematic representation of the transformation of E"(Cu) Center (a) into E"(Cu) Silent Center (b).

We believe that the process of σ bonds formation is driven by the accumulation of large amount of water molecules by the material during the exposure to air moisture, until many cavities of the network become completely filled. Since in the pristine structure of HKUST-1 the $Cu^{2+}$ ions are naturally located at the surface of the pores, the large amount of water stored into the cavities originates a significant pressure acting on them, ultimately forcing many $Cu^{2+}$ couples and to establish a σ bond.

By inspection of Figure 7 it turns out that new significant processes take place in the material after exposure times to air comprised between 20 and 50 days. They represent the second stage of the process of decomposition of HKUST-1. The main effect observed in this stage is the replacement of the residual E"(Cu) centers with E'$_1$(Cu) centers. From a general point of view, it indicates that a second distinguishable structural relaxation process affects the residual E"(Cu) which have not disappeared in the first stage (0 < τ < 20 days). Interestingly, this new process transforms each residual S=1 center into a S=1/2 center, putting forward that just one of the two copper ions involved in those triplet centers actually becomes EPR silent. In order to explain this change of the magnetic properties of the material we suppose that a change of the oxidation state takes place in one of the two ions involved in the related paddle-wheels, reasonably from $Cu^{2+}$ to $Cu^{1+}$, suggesting that some relevant structural changes are involved in this case. In line with this expectation, data obtained from BET analysis show that about 50% of the pristine porosity of the material is irreversibly lost after 36 days of exposure to air (Table 1). Accordingly, we have found a significant reduction of the amplitude of the XRD peaks (Figure 3) and a deterioration of the quality of the material grains surfaces, pointed out by SEM microscopy (Figure 4c). The occurrence of irreversible processes is also confirmed by EPR data, which show that the pristine concentration of E"(Cu) centers is not recovered upon reactivation of the sample (data not shown). In order to take into account all these experimental evidences we suggest that the hydrolysis of the Cu-O bonds takes place in this second stage of the process of decomposition of HKUST-1. In fact, it is known from studies performed by IR and Raman spectroscopy[34] that when water reaches the liquid phase into the pores of the material it becomes able to hydrolyze the Cu-O bonds. Furthermore, since it is well shown that a missing ligand in a paddle-wheel causes the reduction of the oxidation state from $Cu^{2+}$ to $Cu^{1+}$ of one of the metal ions,[59] here we suppose that, as a consequence of the hydrolysis process, a carboxylate bridge is detached from the paddle-wheel, as shown in Figure 9.

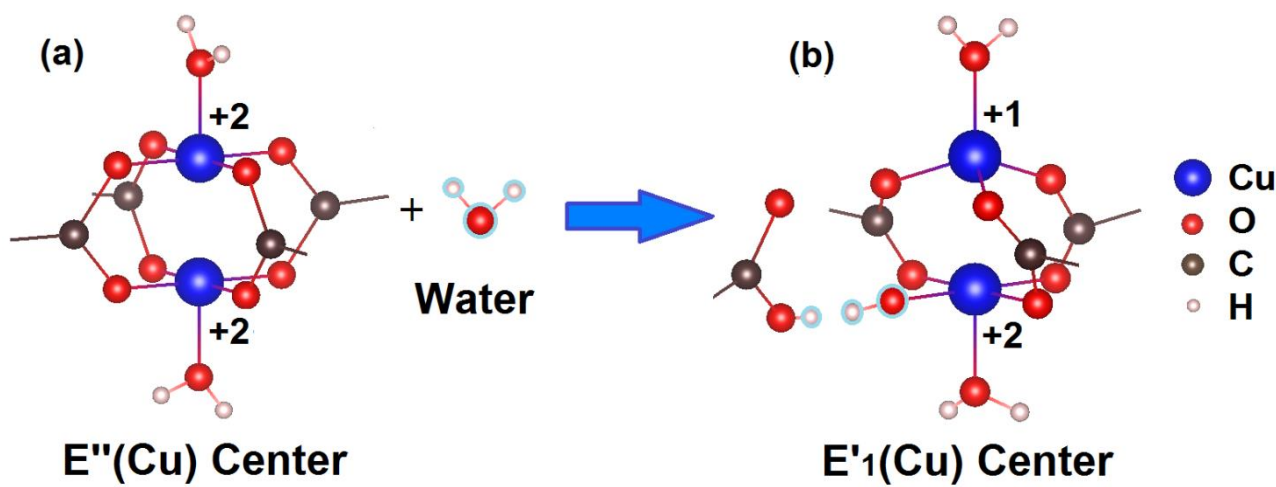

**Figure 9:** Schematic representation of E"(Cu) Center (a) interacting with water molecule, which promotes the hydrolysis of a Cu-O bond, the detaching of a carboxylate bridge and the formation of E'$_1$(Cu) Center (b).

As shown, the copper ion directly affected by the hydrolysis coordinates with OH$^-$ group in replace to the carboxylate bridge (lower ion in Figure 9b), whereas the latter ion simply reduces its initial coordination number from five to four and stabilizes in an almost tetrahedral geometry (upper ion in Figure 9b). This latter geometry is presumably mainly driven by the carboxylate bridge facing the detached one. In fact, its strain becomes unbalanced after the hydrolysis process and it consequently forces the three remaining basal Cu-O bonds to equal their mutual angles (upper ion in Figure 9b). The tetrahedral geometry is also well known to favor the change of the oxidation state of the copper ion from $Cu^{2+}$ to $Cu^{1+}$,[60] in excellent agreement with the indications obtained from EPR data and discussed above. The presence of $Cu^{1+}$ ions in HKUST-1 is not quite surprising. In fact, as anticipated, it has been already observed in samples in which one linker is missing in the paddle-wheel[59] but also in HKUST-1 subjected to thermal treatments[60-61] or soft X-rays irradiation.[62] A property of the second stage of the decomposition process of HKUST-1 upon exposure to air moisture, which needs to be underlined, consists in the fact that our data indicate that the hydrolysis is unable to affect the paddle-wheels involved in the first stage of the process, i.e. those in which the two nearby copper ions have established a σ bond. This property, is very interesting because it proves that in appropriate conditions the paddle-wheels of the material are able to relax reversibly towards a structure which is unaffected by interaction with moisture for times of at least six month.

By inspection of Figure 9b, it is easy to recognize that there is no full charge compensation around the Cu ions involved in the $E'_1(Cu)$ center. In fact, since each oxygen of the carboxylate bridges and each $OH^-$ group brings a nominal charge of ½ $e^-$ and $e^-$, respectively, an overall excess of charge of ½ $e^-$ results on both the copper ions. Furthermore, it is reasonable to expect that the structure of the $E'_1(Cu)$ centers also suffers of a relevant local strain. Indeed, the detaching of one carboxylate bridges from the paddle-wheels, driven by the hydrolysis of the Cu-O bonds, significantly reduces the pristine symmetry of the paddle-wheels, compromising the natural compensation of strains taking place when the carboxylate bridges are arranged in couples, facing one another and pulling in opposite directions. This unreleased stress is indeed revealed by the absence of the hyperfine quadruplet characteristic of S=1/2 copper centers[20,39-40] in the EPR spectrum of $E'_1(Cu)$ at 77 K (Figure 6). In fact, strain increases the natural inhomogeneous distribution of the paramagnetic centers and makes the usually poorly resolved peaks of the quadruplet to become essentially unrecognizable.[39-40]

Such overall strain induced by the large concentration of E'$_1$(Cu) centers (~ 4x10$^{20}$ spin/cm$^3$) during the second stage of the decomposition process should also influence measurably the global properties of the material. In good agreement with this conjecture, the XRD peaks observed for HKUST-1 exposed to air moisture for 36 days are significantly shifted toward smaller angles with respect those observed in the fresh sample, revealing a considerable swelling of the matrix (Figure 3).

The lack of charge balance and the presence of unreleased stress in the E'$_1$(Cu) center are not surprising, as the latter is just a structural defect generated by the hydrolysis of Cu-O bonds and not a structure stabilized during the synthesis process. However, these features suggest that if more stable structures are accessible, in which the complete charge balance can be established and the stress can be released, then they will be actually reached in a time dictated by the thermodynamic properties of the system. Indeed, the data reported in Figure 7 suggest that the expected relaxation process of the metastable E'$_1$(Cu) centers towards more stable structures is actually observed experimentally starting from exposure times of the order of 50 days. It constitutes the third stage of the process of decomposition of the HKUST-1 upon exposure to moisture. The details of the relaxation processes involved in this stage can be easily unveiled by remembering that each carboxylate bridge chelates the two copper ions of the paddle-wheels and it contributes to the negative charge surrounding each ion by about ½ e$^-$. Since this latter value actually corresponds to the charge unbalance suffered by the E'$_1$(Cu) center, we expect that the system will restore the charge balance by simply detaching a second carboxylate bridge from the same structure. Considering that the geometrical disposition of the three carboxylate bridges involved in the E'$_1$(Cu) center is not equivalent, two distinguishable structures can originate from the second detaching event. These two structures are shown in Figures 10b and 10c.

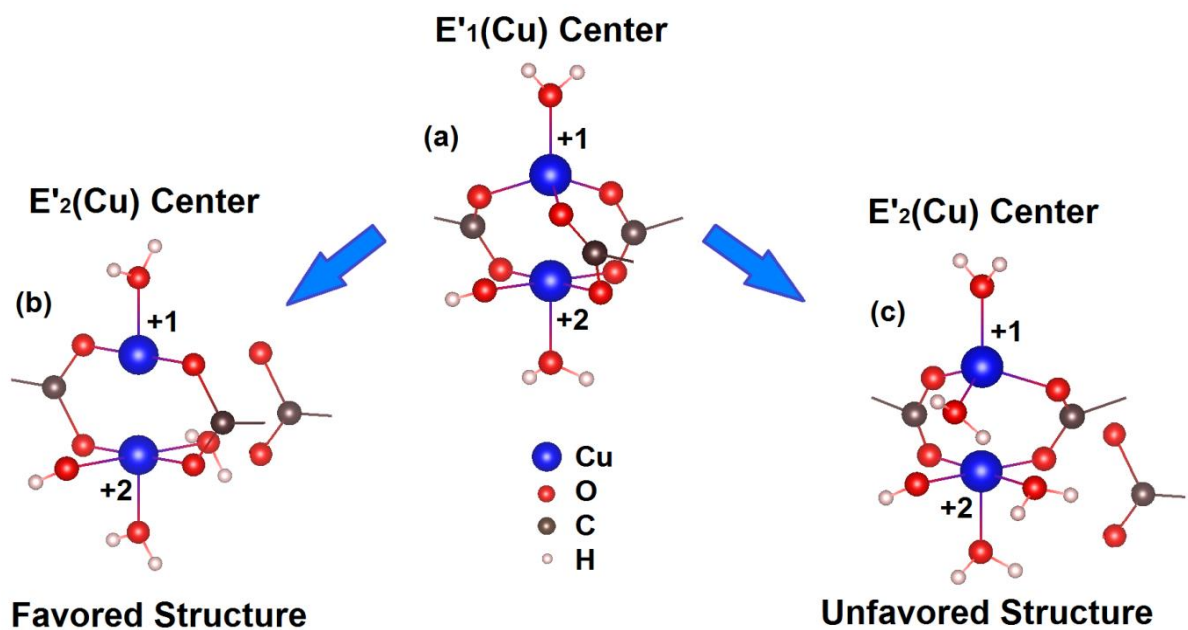

**Figure 10:** Schematic representation of E'$_1$(Cu) Center (a) which relaxes into E'$_2$(Cu) Center with two possible structures favored (b) and unfavored (c).

The first structure (Figure 10b) originates when the two detached carboxylate bridges were facing one another. In this case the second detaching process leaves the Cu$^{1+}$ ion in a T-shaped structure. Since this geometry is well compatible with the oxidation state +1 for copper,[63-64] we assume that no other structural changes take place in this site. At variance, the second copper ion involved in the same structure (Figure 10b) presumably coordinates to a water molecule in place of the carboxylate bridge after detaching, so that it recovers the almost square pyramidal geometry which is suitable for Cu$^{2+}$ ions. The second structure (Figure 10c) originates when the two detached carboxylate bridges were side by side. In this case both the copper ions involved in the structure presumably coordinate a water molecule in place of the carboxylate bridge after the second detaching, so preserving the same geometry as before the last detaching. From the magnetic point of view, the two structures of Figures 10b and 10c are S=1/2 species and in principle both of them may contribute to the EPR signal we attribute to the E'$_2$(Cu) center. However, we believe that the former structure (Figure 10b) is strongly favored with respect to the latter (Figure 10c), as it involves just two facing carboxylate bridges mutually balancing their strains. Consistently with this assumption, the characteristic hyperfine quadruplet related to the copper nucleus is easily recognizable in the EPR line shape of E'$_2$(Cu) center (Figure 6), indicating that the strain affecting the E'$_1$(Cu) center is actually released when E'$_1$(Cu) relaxes towards E'$_2$(Cu). This picture is actually confirmed by quantitative analysis of the spectra. In fact, computer simulations of the EPR lines attributed to E'$_1$(Cu) and E'$_2$(Cu) clearly indicate a strong reduction of the strain associated to $g_3$ going from the former to the latter structure (Table 2). The occurrence of strain release also reflects itself on the global properties of the crystalline lattice, as indicated by XRD measurements (Figure 3), which show that in the third stage of the process of decomposition of HKUST-1 the peaks recover the position they naturally occupy in the unbroken pristine structure of the material. The gradual reduction of XRD peaks intensity recognizable in Figure 3 upon exposure of HKUST-1 to air is reasonable attributable to the loss of crystallinity resulting from the first and the second detaching processes.

As a final remark, it is worth noting that the concentration of E'$_2$(Cu) centers generated for exposure times larger than 50 days is lower by about 30% with respect to that of preexisting E'$_1$(Cu) centers. This result suggests that another relaxation channel is possible for the E'$_1$(Cu) center, which induces a change of the oxidation state of the paramagnetic copper ion involved in the defect from $Cu^{2+}$ to $Cu^{1+}$ and finally generate a full EPR silent hydrolyzed paddle-wheel. Since this latter minor specie is not observable by EPR, its atomic scale structure cannot be further investigated in the present experimental investigation.

The models proposed above to describe the third stage of the process of decomposition of HKUST-1 upon exposure to air agree very well also with BET and SEM data. In particular, these data show that the degree of decomposition of the material in the third stage of the process is comparable with respect to that observed in the second stage, further confirming that the paddle-wheels involved in the first stage of the process are not involved at all in the decomposition of the material dictated by the hydrolysis process.

Rigorously, the comprehensive three-stages process of decomposition described above concerns just the effects induced by the first prolonged exposure of fresh HKUST-1 to air moisture. Our work has been focused on this first exposure because it is the most important one, as it compromises the integrity of the pristine crystalline network and because it produces the largest effects of decomposition in the material. Nevertheless, in many applications, samples of HKUST-1 are actually subjected to repeated processes of activations and subsequent exposures to moisture. It is well known, from previous studies, that after each of these cycles a further fraction of the material decomposes irreversibly, suggesting that processes similar to those described above takes place in cyclic way, affecting the fraction of the material escaped to the decomposition processes induced by previous exposures.

To conclude, we would like to note that our choice to expose HKUST-1 to air at just 70% RH and 300 K is obviously related to the fact these conditions very well mimic those commonly encountered in ambient conditions. However, we have found indications that the picture we have found here of the process of decomposition of HKUST-1 actually is quite general. In fact, many our test experiments performed by exposing HKUST-1 to air with different values of RH and temperatures have shown that all the processes described above actually takes place also in these latter cases, but just after different characteristic times of exposure to the specific environment with respect to those obtained in the present work. The generality of the picture we have drawn here of the process of decomposition of HKUST-1 is also supported by the fact that, as discussed above, EPR spectra similar to those we report in Figure 5 for the *SAMPLE B* have been already reported in previous works by other authors for HKUST-1 samples exposed to moisture in significantly different conditions.[21]

CONCLUSIONS

The experimental results presented here show that many complex structural processes take place when HKUST-1 interacts with air moisture for long time. All the models proposed above will be helpful for possible future works aimed at verifying these conclusions by quantum chemical calculations. Nevertheless, thanks to the appropriate choice of the set of experimental techniques considered in the present work, many fundamental properties of these processes are unveiled. First of all, we have recognized for the first time that the process of decomposition of HKUST-1 takes place through three different stages. All the processes affecting the network of the material during these stages are described in great detail at the atomic scale level. In particular, we have found that during the first stage a reversible process takes place, which makes about 65% of the couples of nearby $Cu^{2+}$ ions of the material to establish a σ bond. For longer exposure times, during the second and the third stages of the process, also the remaining 35% of the $Cu^{2+}$ ions couples are affected by interaction with moisture. However, in these latter cases the processes involved are clearly irreversible. They involve the hydrolysis of the Cu-O bonds and generate two characteristic paramagnetic centers, here characterized and named $E'_1(Cu)$ and $E'_2(Cu)$, respectively. Interestingly, our results indicate that hydrolysis is unable to affect the copper ions involved in the first stage of the process. This result is particularly important as it indicates that it is possible to make the HKUST-1 very stable with respect to interaction with moisture by forcing the couples of $Cu^{2+}$ ions involved in the paddle-wheels to establish a σ bond. We believe that this new idea could inspire future works devoted to the identification of specific procedures able to stabilize the HKUST-1 towards moisture, allowing easiest and cheapest ways to store this outstanding material in ordinary ambient conditions.

## ASSOCIATED CONTENT

**Supporting Information**

In supporting Information we report further details about experimental EPR spectra acquired at 300 K, isotherms of $N_2$ adsorption, comparison between simulated and experimental curves as well as EPR spectra of activated samples of HKUST-1.

## AUTHOR INFORMATION

**Corresponding Author**

*Gianpiero Buscarino

Dipartimento di Fisica e Chimica, Università di Palermo, Via Archirafi 36 - 90123 Palermo, Italia.

e-mail: gianpiero.buscarino@unipa.it

Telephone number +39 091-23891725

## ACKNOWLEDGMENT

The authors thank the people of the LAMP group (www.unipa.it/lamp) at the Department of Physics and Chemistry of the University of Palermo for useful discussions, Dr. Francesco Giordano, CNR-ISMN, Palermo, for the assistance offered during XRD measurements and Mr. Francesco Furnari, Department of DISTEM of the University of Palermo for the support during SEM measurements.

TOC GRAPHIC

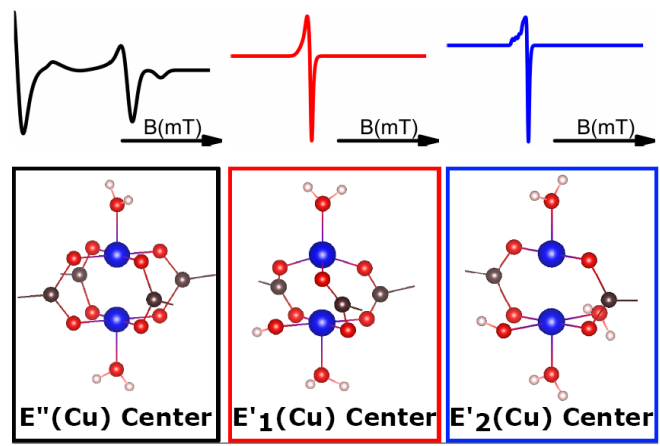